\documentstyle[12pt]{article}

\begin{document}

\title{FERMIONS IN THE BACKGROUND OF THE SPHALERON BARRIER}
\vspace{1.5truecm}
\author{
{\bf Jutta Kunz}\\
Instituut voor Theoretische Fysica, Rijksuniversiteit te Utrecht\\
NL-3508 TA Utrecht, The Netherlands\\
and\\
Fachbereich Physik, Universit\"at Oldenburg, Postfach 2503\\
D-2900 Oldenburg, Germany
\and
{\bf Yves Brihaye}\\
Facult\'e des Sciences, Universit\'e de Mons-Hainaut\\
B-7000 Mons, Belgium}

\vspace{1.5truecm}

\date{January 26, 1993}

\maketitle
\vspace{1.0truecm}

\begin{abstract}
We demonstrate the level crossing phenomenon
for fermions in the background field of the sphaleron barrier,
by numerically determining the fermion eigenvalues along
the minimal energy path from one vacuum to another.
We assume that the fermions of a doublet are degenerate in mass,
allowing for spherically symmetric ans\"atze
for all of the fields, when the mixing angle dependence
is neglected.
\end{abstract}
\vfill\eject

\section{Introduction}

In 1976 't Hooft [1] observed that
the standard model does not absolutely conserve
baryon and lepton number
due to the Adler-Bell-Jackiw anomaly.
The process 't Hooft considered was
spontaneous fermion number violation due to instanton
induced transitions.
Fermion number violating
tunnelling transitions between
topologically distinct vacua might indeed
be observable at high energies at future accelerators [2,3].

The possibility of fermion number
violation in the standard model
was considered from another point of view by
Manton [4].
Investigating the topological structure
of the configuration space of the Weinberg-Salam theory,
Manton showed that there are noncontractible
loops in configuration space, and
predicted the existence of a static, unstable solution
of the field equations,
a sphaleron [5], representing
the top of the energy barrier between
topologically distinct vacua.

At finite temperature this energy barrier between
topologically distinct vacua can be overcome
due to thermal fluctuations of the fields,
and fermion number violating
vacuum to vacuum transitions
involving changes of baryon and lepton number
can occur.
The rate for such baryon number violating processes
is largely determined by a Boltzmann factor,
containing the height of the barrier at a given
temperature and thus the energy of the sphaleron.
Baryon number violation in the standard model due to
such transitions over the barrier may be relevant
for the generation of the baryon asymmetry of the universe
[6-10].

How can baryon and lepton number change
when the barrier between topologically distinct vacua
is traversed?
The answer is seen in the level crossing picture.
Let us consider a process
which starts in the vacuum sector
labelled by the Chern-Simons number $N_{\rm CS}$.
During the process the barrier is traversed.
The Chern-Simons number changes continuously,
ending in the vacuum sector $N_{\rm CS}-1$.
Let us assume a filled Dirac sea
in the first vacuum.
While the gauge and Higgs field configurations slowly change
and with them the Chern-Simons charge,
the fermion levels also change.
When the bosonic configurations reach the top of the barrier,
the sphaleron with Chern-Simons charge $1/2$,
one fermion level of the sea has precisely reached zero energy,
and when the bosonic fields reach
the next vacuum configuration,
this occupied energy level has dived out of the Dirac sea.

In this letter we demonstrate the level crossing phenomenon
for fermions in the background field of the sphaleron barrier,
by numerically determining the fermion eigenvalues along
the minimal energy path from one vacuum to another [11,12].
We assume that the fermions of a doublet are degenerate in mass.
This assumption, violated in the standard model,
allows for spherically symmetric ans\"atze
for all of the fields, when the mixing angle dependence
is neglected (which is an excellent approximation [13,14]).
At the top of the barrier,
in the background field of the sphaleron,
the fermions reach a zero mode [15-17].

We briefly review in section 2
the Weinberg-Salam lagrangian with the approximations employed.
In section 3 we present the sphaleron energy barrier,
providing the background field for the fermions.
In section 4 we derive the radial equations for
the fermions,
and we present our results in section 5.

\section{\bf Weinberg-Salam Lagrangian}

Let us consider the bosonic sector of the Weinberg-Salam theory
in the limit of vanishing mixing angle.
In this limit the U(1) field
decouples and can consistently be set to zero
\begin{equation}
{\cal L}_{\rm b} = -\frac{1}{4} F_{\mu\nu}^a F^{\mu\nu,a}
+ (D_\mu \Phi)^{\dagger} (D^\mu \Phi)
- \lambda (\Phi^{\dagger} \Phi - \frac{v^2}{2} )^2
\   \end{equation}
with the SU(2)$_{\rm L}$ field strength tensor
\begin{equation}
F_{\mu\nu}^a=\partial_\mu V_\nu^a-\partial_\nu V_\mu^a
            + g \epsilon^{abc} V_\mu^b V_\nu^c
\ , \end{equation}
and the covariant derivative for the Higgs field
\begin{equation}
D_{\mu} \Phi = \Bigl(\partial_{\mu}
             -\frac{i}{2}g \tau^a V_{\mu}^a \Bigr)\Phi
\ . \end{equation}
The ${\rm SU(2)_L}$
gauge symmetry is spontaneously broken
due to the non-vanishing vacuum expectation
value $v$ of the Higgs field
\begin{equation}
    \langle \Phi \rangle = \frac{v}{\sqrt2}
    \left( \begin{array}{c} 0\\1  \end{array} \right)
\ , \end{equation}
leading to the boson masses
\begin{equation}
    M_W = M_Z =\frac{1}{2} g v \ , \ \ \ \ \ \
    M_H = v \sqrt{2 \lambda}
\ . \end{equation}
We employ the values $M_W=80 {\rm GeV}$, $g=0.67$.

For vanishing mixing angle,
considering only fermion doublets degenerate in mass,
the fermion lagrangian reads
\begin{eqnarray}
{\cal L}_{\rm f} & = &
   \bar q_{\rm L} i \gamma^\mu D_\mu q_{\rm L}
 + \bar q_{\rm R} i \gamma^\mu \partial_\mu q_{\rm R}
   \nonumber \\
           & - & f^{(q)} \bar q_{\rm L}
	   (\tilde \Phi u_{\rm R} + \Phi d_{\rm R})
               - f^{(q)} (\bar d_{\rm R} \Phi^\dagger
                     +\bar u_{\rm R} \tilde \Phi^\dagger)
	   q_{\rm L}
\ , \end{eqnarray}
where $q_{\rm L}$ denotes the lefthanded doublet
$(u_{\rm L},d_{\rm L})$,
while $q_{\rm R}$ abbreviates the righthanded singlets
$(u_{\rm R},d_{\rm R})$,
with covariant derivative
\begin{equation}
D_\mu q_{\rm L} = \Bigl(\partial_{\mu}
             -\frac{i}{2}g \tau^a V_{\mu}^a \Bigr) q_{\rm L}
\ , \end{equation}
and with $\tilde \Phi = i \tau_2 \Phi^*$.
The fermion mass is given by
\begin{equation}
M_F=\frac{1}{\sqrt{2}}f^{(q)} v
\ . \end{equation}

All gauge field configurations can be classified by a charge,
the Chern-Simons charge.
The Chern-Simons current
\begin{equation}
 K_\mu=\frac{g^2}{16\pi^2}\varepsilon_{\mu\nu\rho\sigma} {\rm Tr}(
 {\cal F}^{\nu\rho}
 {\cal V}^\sigma
 + \frac{2}{3} i g {\cal V}^\nu {\cal V}^\rho {\cal V}^\sigma )
\   \end{equation}
(${\cal F}_{\nu\rho} = 1/2 \tau^i F^i_{\nu\rho}$,
${\cal V}_\sigma = 1/2 \tau^i V^i_\sigma$)
is not conserved,
its divergence $\partial^\mu K_\mu$
represents the U(1) anomaly.
The Chern-Simons charge
of a configuration is given by
\begin{equation}
N_{\rm CS} = \int d^3r K^0
\ . \end{equation}
For the vacua the Chern-Simons charge is identical to the
integer winding number,
while the barriers are characterized
by a half integer Chern-Simons charge.

\section{\bf Sphaleron Energy Barrier}

The height of the barrier can be obtained by
constructing families of field configurations
for the gauge and Higgs fields,
which interpolate smoothly from one vaccum to another
as a function of the Chern-Simons charge.
Each of these families of configurations
has a maximal energy along
such a path. By finding the minimal value of these
maximal energies one has found the height of the
barrier, the sphaleron [4,5].

In the limit of vanishing mixing angle
the general static, spherically symmetric ansatz for the
gauge and Higgs fields is given by [18]
\begin{eqnarray}
    \Phi & = & \frac{v}{\sqrt {2}}
  \Bigl(H(r) + i \vec \tau \cdot \hat r K(r)\Bigr)
    \left( \begin{array}{c} 0\\1  \end{array} \right)
  \ , \\
  V_i^a & = & \frac{1-f_A(r)}{gr} \epsilon_{aij}\hat r_j
  + \frac{f_B(r)}{gr} (\delta_{ia}-\hat r_i \hat r_a)
  + \frac{f_C(r)}{gr} \hat r_i \hat r_a  \ , \\
  V_0^a & = & 0
\ , \end{eqnarray}
and involves the five radial functions $H(r)$, $K(r)$,
$f_A(r)$, $f_B(r)$ and $f_C(r)$.

This ansatz leads to the energy functional
\begin{eqnarray}
 E & = & \frac{4\pi M_W}{g^2} \int^{\infty}_0 dx
         \Bigl[  \frac{1}{2x^2} (f^2_A + f^2_B  -1)^2
         + (f'_A + \frac{f_Bf_C}{x})^2
         + (f'_B - \frac{f_Af_C}{x})^2
\nonumber \\
   & + & (K^2+H^2) (1+f_A^2+f^2_B +
         \frac{f_C^2}{2})+2f_A (K^2-H^2) - 4f_B H K
\nonumber \\
   & + & 2x^2(H'^2+K'^2) - 2xf_C (K'H - KH') +
         \frac{4\lambda}{g^2} x^2 (H^2+K^2 -1)^2
    \Bigr]
\ , \end{eqnarray}
where $x=M_Wr$,
and to the Chern-Simons number
\begin{equation}
N_{\rm CS} = \frac{1}{2\pi} \int^{\infty}_0 \ dx \Bigl[
(f_A^2+f^2_B) (\frac{f_C}{x} - \varphi') -
(\frac{f_C}{x} - \Theta') -
\Bigl(\sqrt {(f_A^2+f^2_B)}\
\sin (\varphi - \Theta)\Bigr)'\ \Bigr]
\   \end{equation}
with $\varphi = {\rm arctan}({f_B}/{f_A})$.
The function $\Theta(x)$ is an arbitrary radial function,
associated with
the residual gauge invariance of the ansatz (11)-(13).
This gauge freedom can be used to eliminate one of the functions.
Here we choose the radial gauge with the
gauge condition $f_C(x)=0$.

Let us now consider families of configurations,
which connect one vacuum ($N_{\rm CS}=0$)
with another vacuum ($N_{\rm CS}=1$)
passing the sphaleron ($N_{\rm CS}=1/2$).
Note, that the Chern-Simons number of the sphaleron
is independent of the Higgs mass,
$N_{\rm CS}=1/2$ [5].
For this purpose we extremize the functional [11,12]
\begin{equation}
 W = E +   \frac{8 \pi M_W}{g^2} \  \xi \ N_{\rm CS}
\ , \end{equation}
where $\xi$ is a lagrange multiplier.
The minimal energy path
constructed accordingly for $M_H=M_W$
is shown in Fig.~1.
This path is symmetric
with respect its top, the sphaleron.
For large values of the Higgs mass additional
less symmetric sphaleron solutions, bisphalerons, appear [19,20].
The first bisphaleron takes over the role of
the sphaleron. It represents the top of an asymmetric
barrier [12,21], having a Chern-Simons number different from $1/2$.

\section{\bf Fermion Equations}

Let us now consider the fermions in the background of
the sphaleron barrier.
To retain spherical symmetry
we consider only fermion doublets degenerate in mass.
{}From the fermion lagrangian (7)
we obtain the eigenvalue equations
for the lefthanded doublet
\begin{equation}
i D_0 q_{\rm L} + i \sigma^i D_i q_{\rm L}
-f^{(q)} (\tilde \Phi u_{\rm R} + \Phi d_{\rm R} )=0
\ , \end{equation}
and for the righthanded singlets
\begin{equation}
i \partial_0
\left( \begin{array}{c} u_{\rm R}\\d_{\rm R} \end{array} \right)
-i \sigma^i \partial_i
\left( \begin{array}{c} u_{\rm R}\\d_{\rm R} \end{array} \right)
-f^{(q)}
\left( \begin{array}{c} \tilde \Phi^\dagger q_{\rm L}\\
	     \Phi^\dagger q_{\rm L} \end{array} \right)
	     =0
\ . \end{equation}

Employing the spherically symmetric ansatz for
the fermion eigenstates, the hedgehog ansatz,
\begin{equation}
q_{\rm L}(\vec r\,,t) = e^{-i\omega t}
\bigl( G_{\rm L}(r)
+ i \vec \sigma \cdot \hat r F_{\rm L}(r) \bigr) \chi_{\rm h}
\ , \end{equation}
\begin{equation}
q_{\rm R}(\vec r\,,t) = e^{-i\omega t}
\bigl( G_{\rm R}(r)
+ i \vec \sigma \cdot \hat r F_{\rm R}(r) \bigr) \chi_{\rm h}
\ , \end{equation}
with the hedgehog spinor
satisfying the spin-isospin relation
$\vec \sigma \chi_{\rm h} + \vec \tau \chi_{\rm h} = 0 $,
we obtain the following set of four
coupled first order differential equations
\begin{equation}
\tilde \omega G_{\rm L} - F'_{\rm L} - \frac{2}{x}F_{\rm L}
+\frac{1-f_A}{x} F_{\rm L}
-\frac{f_B}{x} G_{\rm L}
-\frac{f_C}{2x}G_{\rm L}
-\tilde M_F(H G_{\rm R} + K F_{\rm R}) = 0
\ , \end{equation}
\begin{equation}
\tilde \omega F_{\rm L} + G'_{\rm L}
+\frac{1-f_A}{x} G_{\rm L}
+\frac{f_B}{x} F_{\rm L}
-\frac{f_C}{2x}F_{\rm L}
-\tilde M_F(H F_{\rm R} - K G_{\rm R}) = 0
\ , \end{equation}
\begin{equation}
\tilde \omega G_{\rm R} + F'_{\rm R} + \frac{2}{x}F_{\rm R}
-\tilde M_F(H G_{\rm L} - K F_{\rm L}) = 0
\ , \end{equation}
\begin{equation}
\tilde \omega F_{\rm R} - G'_{\rm R}
-\tilde M_F(H F_{\rm L} + K G_{\rm L}) = 0
\ , \end{equation}
where $x$ is the dimensionless coordinate,
$\tilde \omega$ is the
dimensionless eigenvalue
$\tilde \omega = \omega /M_W$
and $\tilde M_F$ is the dimensionless fermion mass
$\tilde M_F= M_F/M_W$.
(Remember the gauge choice $f_C=0$.)

The eigenvalue problem (21)-(24) for the fermions
in a sphaleron-like background field
requires certain boundary conditions
for the fermion functions.
At the origin $G_{\rm L}(x)$ and $G_{\rm R}(x)$
are finite, while $F_{\rm L}(x)$ and $F_{\rm R}(x)$
vanish, at spatial infinity all functions vanish.

\section{\bf Results}

In the background field of the sphaleron
the fermions have a zero mode,
i.~e.~a normalizable eigenstate
with zero eigenvalue. In this case, the two functions
$F_{\rm L}(x)$ and $F_{\rm R}(x)$ decouple and are identically
equal to zero.
When the mass of the fermions vanishes,
also $G_{\rm R}(x)$ decouples and the zero mode
can be given analytically [15-17].
The normalized eigenfunctions are shown in Fig.~2
for fermion masses of $M_f=80$ GeV, $8$ GeV and $0.8$ GeV.
In the background field of the bisphalerons
the fermions do not have a zero mode,
in fact, the fermion eigenvalue
depends on the Higgs mass [22].

When the fermion eigenvalue equations are solved
in the background field of the sphaleron barrier,
given by the minimal energy path discussed above,
the level crossing phenomenon is observed.
Since the barrier is
symmetric about the sphaleron, the fermion eigenvalue
$\omega$ is antisymmetric with respect to
the sphaleron configuration.
In Fig.~3 we represent the fermion eigenvalue along the barrier
for fermion masses of $M_f=80$ GeV, $8$ GeV and $0.8$ GeV.
We observe that light fermions are bound only close to the top
of the barrier,
the sphaleron configuration, while heavy fermions are bound
almost along the full path along the barrier.
Denoting by $M_f^{\rm cr}$ the fermion mass, at which
for a given Chern-Simons number, the fermion bound
state enters the continuum,
this observation is illustrated also in Fig.~4, where
the critical fermion mass $M_f^{\rm cr}$
is shown as a function of the Chern-Simons number $N_{\rm CS}$.
At zero mass, fermions are bound only by the sphaleron.

\vfill\eject

\section{Figure Captions}

\noindent{\bf Figure 1}

 The Energy $E$ in TeV is shown
 as a function of the Chern-Simons number $N_{\rm CS}$
 along the minimal energy path
 from one vacuum to another vacuum for $M_H=M_W$.
\vskip 0.3cm

\noindent{\bf Figure 2}

 The fermion zero mode wavefunction components
 $G_{\rm L}(x)$ (positive) and $G_{\rm R}(x)$ (negative) are shown
 for fermion masses $M_f=80$ GeV (solid),
 $M_f=8$ GeV (dashed), and $M_f=800$ MeV (dotted)
 for $M_H=M_W$.
\vskip 0.3cm

\noindent{\bf Figure 3}

 The normalized fermion eigenvalue $\omega/M_f$
 is shown as a function of the Chern-Simons number $N_{\rm CS}$
 along the minimal energy path
 from one vacuum to another vacuum for $M_H=M_W$
 for the fermions masses $M_f=80$ GeV, $M_f=8$ GeV and $M_f=800$ MeV.
\vskip 0.3cm

\noindent{\bf Figure 4}

The critical fermion mass $M_f^{\rm cr}$ (in GeV) at which
the bound state enters the continuum,
for a given Chern-Simons number,
is shown as a function of the Chern-Simons number $N_{\rm CS}$
for $M_H=M_W$.
\vfill\eject

\end{document}